# Pressure-Induced Reversal of Thermal Anisotropy in $Bi_2O_2Se$

Zunyi Deng,[1] Wenwen Xuan,[2] Bin Wei,[2,*] Yongheng Li[3,*]

[1]Engineering Research Center of Integrated Circuit Packaging and Testing, Ministry of Education, Gansu Integrated Circuit Packaging and Testing Industry Research Institute, Tianshui Normal University, Tianshui 741001, China

[2] School of Materials Science and Engineering, Henan Polytechnic University, Jiaozuo 454000, China

[3] Zhongguancun Academy, Haidian, Beijing 100195 China

[*]Corresponding authors emails:

*Bin Wei (binwei@hpu.edu.cn),*

*Yongheng Li (liyongheng@zgci.ac.cn).*

**Abstract**

$Bi_2O_2Se$ is an emerging semiconductor with intrinsically low thermal conductivity, making it a promising material for thermoelectric applications. Hydrostatic pressure can effectively tunes the thermal conductivity, with various pressure-dependent trends reported. However, its impact on thermal anisotropy, particularly in the highly anisotropic $Bi_2O_2Se$, remains poorly understood. Here, we report a pressure-driven reversal of thermal anisotropy: $\kappa_z < \kappa_x$ at 0 GPa transforms into $\kappa_z > \kappa_x$ at 60 GPa without phase transition. This stems from distinct phonon dispersions along the $x$- and $z$-directions under pressure, leading to a reshaped group velocity landscape. Below 10 meV, $v_z > v_x$ at both pressures, with a much greater advantage at 60 GPa. Above 10 meV, $v_x > v_z$ at 0 GPa; however, the difference nearly vanishes at 60 GPa. These changes result from anisotropic lattice compression, with the $z$-axis shrinking more significantly than the $x$-axis and suppressing the lone pair activity of Bi atoms. This study calls for revisiting the pressure dependence of thermal conductivity anisotropy and provides insights for pressure-driven thermal switching applications.

Bi$_2$O$_2$Se has also been recognized as a promising thermoelectric oxide because of its low thermal conductivity and favorable electronic properties, offering advantages such as low toxicity and excellent thermal stability. Its thermal conductivity ($\kappa$) is as low as ~1.1 W m$^{-1}$ K$^{-1}$, accompanied by a high Seebeck coefficient of ~500 $\mu$V K$^{-1}$.[1,2] At ambient pressure and room temperature, Bi$_2$O$_2$Se exhibits a thermal conductivity anisotropy ($\kappa_x/\kappa_z$) of nearly 2, which is smaller than that of Bi$_2$Se$_3$ owing to the stronger interlayer electrostatic interactions in Bi$_2$O$_2$Se.[3] It is unlike the weak van der Waals coupling characteristics of conventional layered materials such as Bi$_2$Se$_3$.[3] Tailoring the thermal conductivity of Bi$_2$O$_2$Se is crucial for enhancing its thermoelectric efficiency (TE). Various strategies, including vacancy engineering, strain, doping, and mixed chalcogen networks, effectively enhance the thermoelectric performance of Bi$_2$O$_2$Se.[4–7] For instance, bismuth vacancies improve electronic properties while reducing electronic thermal conductivity, synergistically boosting the figure of merit.[1] Guo et al. predicted that a uniform in-plane strain can triple thermoelectric efficiency by suppressing lattice thermal conductivity.[5] In addition, hydrostatic pressure provides another effective route to modulate thermal transport. Although previous work has shown that pressure can effectively modulate the properties of Bi$_2$O$_2$Se,[8] a detailed understanding of its thermal conductivity under pressure remains lacking.

Recent studies under pressure have shown that thermal conductivity can exhibit diverse pressure-dependent trends, underscoring the complexity of the underlying mechanisms.[9] Conventionally, hydrostatic pressure is expected to increase $\kappa$ due to lattice compression and stronger interatomic interactions, in line with the Liebfried-Schlömann (L-S) theory.[10] On the other hand, thermal conductivity under pressure can decrease or peak and then drop, deviating from conventional expectations. Such behavior may stem from anomalous phonon softening, such as in the transverse acoustic (TA) branches of CuInTe$_2$[11,12] and PdS[13], or strong four-phonon scattering, as observed in BAs[14]. However, most studies have reported thermal conductivity along only one direction under hydrostatic pressure. For anisotropic materials, the pressure dependence of the thermal conductivity anisotropy is equally important. Because hydrostatic pressure compresses the lattice uniformly, anisotropy is expected to remain close to its ambient value before phase transition. In fact, pressure can dramatically reshape thermal anisotropy. For example, *hcp*-Fe and $MgCO_3$ can exhibit enhanced anisotropy under pressure,[15,16] while layered *γ*-InSe shows the opposite trend: its in-plane thermal conductivity decreases while out-of-plane conductivity increases, reducing the anisotropy ratio from 6.95 at 0 GPa to 1.26 at 8 GPa—indicating a shift toward isotropy.[17] More intriguingly, anisotropy inversion, where the originally high-conductivity direction becomes the lower one under pressure, remains largely unexplored. If achievable, it could enable pressure-driven control of the heat flow, offering a route to solid-state thermal switches. $Bi_2O_2Se$, with its pronounced thermal conductivity anisotropy and special layered structure, may fulfill this requirement.

In this study, we systematically investigate the pressure-dependent phonon dispersions and lattice thermal conductivity of $Bi_2O_2Se$ using first-principles calculations and the phonon Boltzmann transport equation. We found that hydrostatic pressure induces a reversal of thermal conductivity anisotropy without phase transition: while $κ_x > κ_z$ at 0 GPa, $κ_z$ surpasses $κ_x$ at high pressure. This inversion stems from anisotropic lattice compression, particularly the reduction of the *z*-axis faster than that of the *x*-axis, which reshapes phonon dispersions. The difference in the phonon group velocities is the primary origin of the anisotropy inversion. At 0 GPa, phonons above ~10 meV exhibit significantly higher group velocities along the *x*-direction than along *the z*-direction, whereas below 10 meV, $v_z$ exceeds $v_x$. In contrast, at 60 GPa, the velocities above 10 meV became nearly comparable, and the advantage of $v_z$ over $v_x$ below 10 meV was dramatically enhanced compared to that at 0 GPa. This low-energy phonon behavior drives a faster increase in $κ_z$ than in $κ_x$ under hydrostatic pressure, ultimately leading to the observed inversion of thermal conductivity anisotropy. Our results challenge the conventional view that hydrostatic pressure universally suppresses thermal anisotropy, highlighting its potential to engineer and even invert anisotropic heat transport.

The plane-wave pseudopotential approach was applied to conduct first-principles calculations using the Vienna ab initio simulation package (VASP).[18] The exchange–correlation interactions were described within the generalized gradient approximation (GGA). A plane-wave kinetic energy cutoff of 550 eV was employed, and the electronic self-consistent loop converged to an energy tolerance of $10^{-6}$ eV. All the atoms in the unit cell were fully relaxed until the residual forces were less than $10^{-4}$ eV/Å. Brillouin-zone integration was performed using a converged $8 \times 8 \times 4$ Monkhorst–Pack[19] k-point mesh. Harmonic and anharmonic interatomic force constants (IFCs) required to solve the phonon Boltzmann transport equation were extracted via the canonical configuration method in the TDEP code.[20] Specifically, we carried out six iterative steps, introducing 4, 8, 16, 32, 64, and 128 new atomic configurations in each step, resulting in a total of 252 structures to obtain converged IFCs. These configurations were generated from conventional cells optimized at 0, 20, 40, and 60 GPa, which were subsequently supercell-expanded to $4 \times 4 \times 2$ to acquire the atomic forces and displacements datasets. The IFCs obtained from TDEP were then converted into the input format required by the Fourphonon code.[21,22] The lattice thermal conductivity was calculated on a dense $12 \times 12 \times 12$ q-point grid. Four-phonon scattering processes are included via sampling acceleration methods, with convergence achieved at 7,000 sampling points.[23]

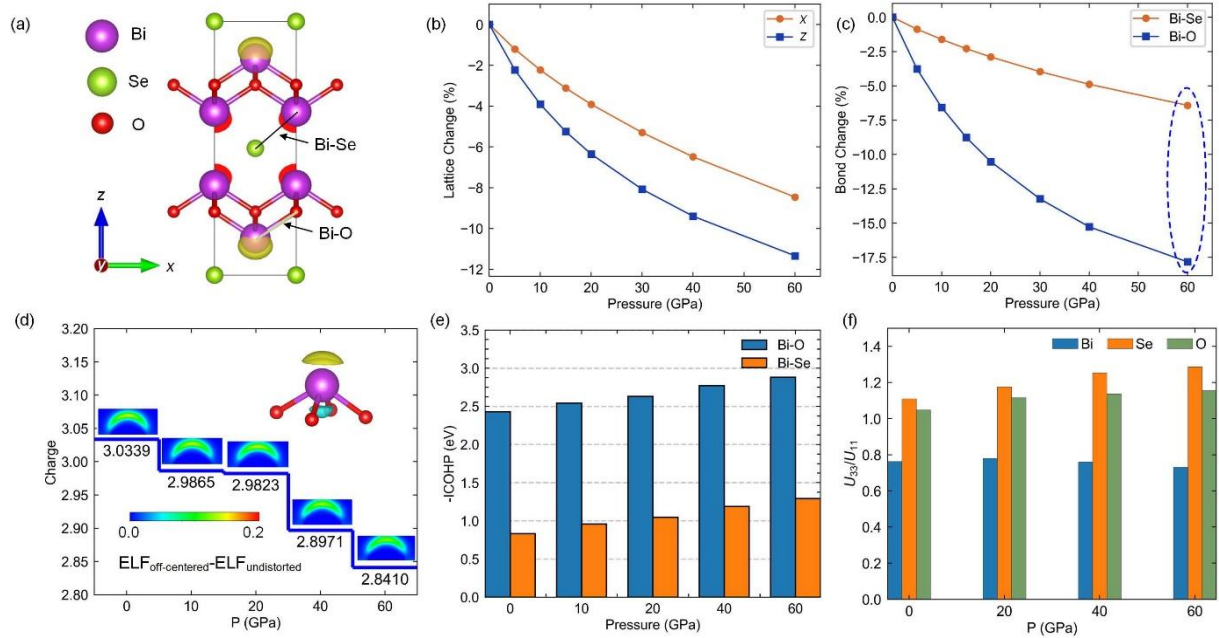

Figure 1. (a) Layered crystal structure of the $Bi_2O_2Se$ conventional cell at 0 GPa, viewed along the z-axis using VESTA [24]. The bismuth atoms show lone pair electron. (b) Pressure dependence of the relative change in the lattice parameters. (c) Evolution of Bi-Se and Bi-O bond lengths under pressure. (d) Variation of the Bi lone pair character with increasing pressure. (e) Pressure-induced changes in the COHP of Bi-O and Bi-Se bonds. (f) Pressure dependence of $U_{33}/U_{11}$ ratio, reflecting anisotropic atomic displacement parameters.

We began by exploring the structural evolution of $Bi_2O_2Se$ under pressure, focusing on the lattice parameters and chemical bonding (Figs. 1a, 1b, and 1c). $Bi_2O_2Se$ possesses a layered structure that belongs to the *I4/mmm* space group. Its $Bi_2O_2$ and Se layers are held together by weak electrostatic interactions rather than van der Waals interactions in $Bi_2Se_3$.[3] Under increasing hydrostatic pressure, the lattice is compressed anisotropically, where the lattice parameter along *z*-axis becomes small faster than that along the *x*-axis, leading to distinct compression rates along different crystallographic directions. This anisotropic lattice response is also reflected in the pressure dependence of the bond lengths. Under compression, the Bi–Se bond shortens significantly slower than the Bi–O bond. At 60 GPa, Bi-Se contracts by only 6.42%, whereas Bi-O shortens by 17.82%, a difference of 11.40%.

We then analyzed how the electronic structure of $Bi_2O_2Se$ evolves under pressure. The Bader charge analysis (Fig. 1d) reveals that under increasing hydrostatic pressure, the charge transfer from Bi to neighboring O atoms gradually decreases, resulting from the progressive suppression of Bi's lone pair under pressure. The O–Bi–O bond angle narrows from 113.44° at 0 GPa to 109.79° at 60 GPa. This behavior contrasts with that of BiCuSeO under tensile strain, where the charge redistribution enhances the lone pair electron (LPE) effect and expands the Bi electron cloud.[24] However, for $Bi_2O_2Se$ under hydrostatic compression, the Bi-centered electron density became increasingly localized, and the LPE was gradually quenched. Electron localization function (ELF) analyses support this trend, compared to the off-center and centrosymmetric Bi configurations. The ELF feature around Bi weakened under pressure (green crescent in Fig. 1d), further visualizing the suppression of the lone pair. The observed suppression of the LPE effect implies pronounced weakening of the lattice anharmonicity in $Bi_2O_2Se$ under hydrostatic pressure.

Figure 1e shows that the integrated crystal orbital Hamilton population (ICOHP) of Bi-O bonds in $Bi_2O_2Se$ increases steadily under hydrostatic pressure, rising by ~0.45 eV from 0 to 60 GPa, as well as the Bi-Se bond. This trend arises because pressure-induced bond shortening enhances orbital overlap, thereby strengthening the chemical bonds. According to the relation $\kappa_L \propto (F/M)^{1/2}$ (where *F* is the bond force constant and *M* is the atomic mass), stronger bonds typically lead to higher lattice thermal conductivity, which is consistent with the general expectation that $\kappa$ increases under pressure. However, the anisotropic structural response implies a direction-dependent enhancement of thermal conductivity. As shown in Fig. 1f, the $U_{33}/U_{11}$ ratio slightly decreases for Bi but clearly increases for Se and O, which maybe result from Bi's lone pair electron changes under pressure. This anisotropic displacement confirms direction-dependent structural changes under pressure, which may lead to distinct modifications in phonon dispersions along different crystallographic directions.

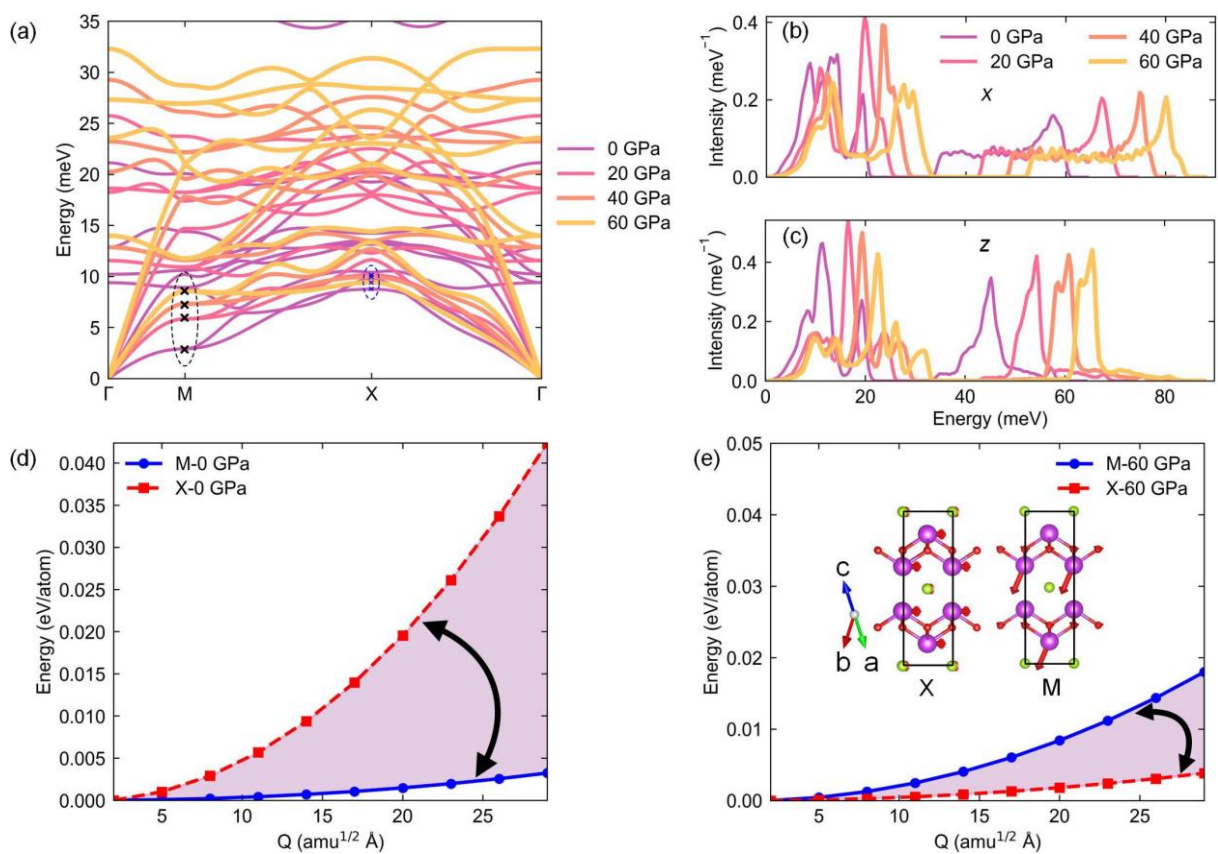

Figure 2. (a) Evolution of phonon dispersion under pressure. The black and blue crosses denote the pressure-dependent frequencies of the TA branch at the M ([0, 0, 1]) and X ([0.5, 0.5, 0]) high-symmetry points, respectively, in the conventional cell. The Γ–M and Γ–X directions correspond to the $z$- and $x$-axes in real space, respectively. (b, c) Phonon density of states projected onto the $x$- and $z$-directions as a function of the pressure. Frozen-phonon potential energy surfaces for the TA mode at M and X at (d) 0 GPa and (e) 60 GPa. The inset in (e) illustrates the atomic displacement patterns of the TA modes at these two points, where the arrow lengths indicate the relative magnitudes of atomic displacements.

Based on the pressure-induced structural changes in $Bi_2O_2Se$, its phonon response was analyzed. As shown in Fig. 2a, most phonon branches harden with pressure. No imaginary frequencies are observed up to 60 GPa, indicating the absence of a structural phase transition. The TA mode at M stiffens significantly, whereas that at X exhibits a weak and nonmonotonic change. This anisotropic phonon evolution of the TA branch likely drives the reversal of the thermal conductivity, where $\kappa_z$ surpasses $\kappa_x$ at high pressures. Figures 2b and 2c show that the phonon density of states (DOS) projected onto both the *x*- and *z*-directions harden significantly with pressure. However, a key difference emerges at low energies. Below ~20 meV, the *x*-projected DOS retain its spectral weight with no notable peak weakening. In contrast, the *z*-projected DOS near 10 meV not only hardens, but also exhibits a marked reduction in intensity, indicating that phonon modes previously clustered around 20 meV disperse into a broader energy range. The weakened low-energy phonon DOS peak, reminiscent of the suppression of rattling-like characteristics, suppresses phonon–phonon scattering along the *z*-direction, thereby may enhance $\kappa_z$ more effectively than $\kappa_x$.

To understand the distinct pressure responses of the TA modes at X and M, we performed frozen-phonon potential analysis. As shown in Figs. 2d and 2e, while the energy difference between the two points decreases under pressure, their behavior diverges: the potential at M increases monotonically (consistent with normal stiffening), whereas that at X exhibits anomalous softening, suggesting the nonmonotonic frequency evolution of the X-point TA mode noted earlier. Notably, softening at X is more pronounced than that at M, suggesting that this anomalous response, driven by changes in lattice anharmonicity, suppresses the increase of thermal conductivity along the *x*-direction relative to that along the *z*-direction. This contrasting phonon behavior may be a kind of factor driving the eventual crossover, where $\kappa_z$ exceeds $\kappa_x$.

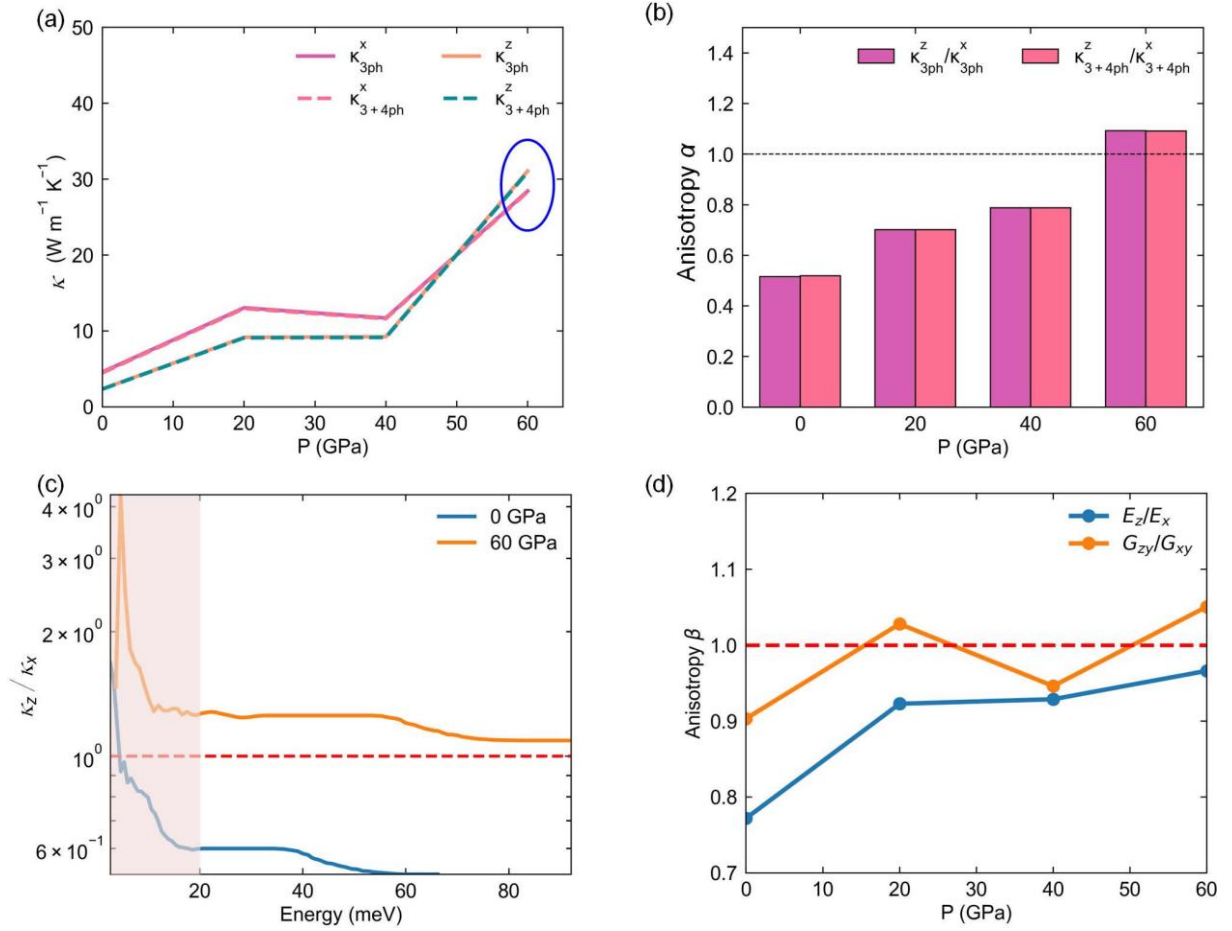

Figure 3. (a) Lattice thermal conductivity ($\kappa$) of $Bi_2O_2Se$ at 300 K as a function of hydrostatic pressure. The subscripts "3ph" and "3+4ph" denote calculations including three-phonon scattering only, and both three- and four-phonon scattering, respectively. (b) Pressure dependence of the $\kappa_z/\kappa_x$ ratio. (c) Ratio of cumulative $\kappa_z$ to $\kappa_x$ as a function of phonon energy at 0 and 60 GPa. (d) Pressure evolution of the ratio of Young's moduli (shear moduli) along the $z$- and $x$-directions.

The lattice thermal conductivity of $Bi_2O_2Se$ at 300 K under pressure is shown in Fig. 3a. At 0 GPa and 300 K, our calculated $\kappa$ for $Bi_2O_2Se$ is slightly higher than the reported ~1.1–2 W m$^{-1}$ K$^{-1}$,[1,3] likely due to differences in force-constant methodology. The $\kappa$ value for 3+4ph is nearly identical to the 3ph-only results, suggesting that the effect of four-phonon scattering is negligible. The calculated wave-like phonon contribution to thermal conductivity by wigner thermal transport equation remains nearly constant at ~0.2 W m$^{-1}$ K$^{-1}$ along $x$ direction and ~0.1 W m$^{-1}$ K$^{-1}$ along $z$ direction with increasing pressure, indicating a negligible effect on the overall trend. Notably, $\kappa_z$ is smaller than $\kappa_x$ below 40 GPa, but surpasses $\kappa_x$ above 50 GPa, revealing a pressure-induced reversal of thermal anisotropy. Figure 3b clearly illustrates the anisotropy reversal: the $\kappa_z/\kappa_x$ ratio is below 1 at 0-40 GPa but exceeds 1 at 60 GPa. The ratio of cumulative $\kappa_z$ to $\kappa_x$ analysis (Fig. 3c) shows that phonons below ~20 meV dominate thermal transport. At 0 GPa, $\kappa_z$ briefly exceeds $\kappa_x$ in a narrow low-energy window, but $\kappa_x$ rises rapidly with increasing phonon energy, driving the ratio $\kappa_z/\kappa_x$ below 1—resulting from the higher group velocity of the Γ–X (in-plane) phonon modes. TA branch compared to that of the Γ-M (out-of-plane) counterpart. However, At 60 GPa, the Γ–M TA mode hardens dramatically and becomes nearly degenerate in frequency with the Γ–X TA mode. However, the X-point potential exhibits anomalous softening (Fig. 2e), signaling enhanced anharmonicity along the in-plane direction. This suppresses the growth of $\kappa_x$ relative to $\kappa_z$, resulting in $\kappa_z/\kappa_x$ larger than 1 across the entire phonon energy range. This reversal is also observed for the mechanical properties (Fig. 3d). The ratio of Young's moduli ($E_z/E_x$) approaches 1, indicating that the LA modes stiffen similarly along both directions with increasing pressure. In contrast, the shear modulus ratio ($G_z/G_x$) exhibits an increasing trend that exceeds 1 under pressure, matching the different behaviors of the TA modes along the $x$ and $z$ directions. This difference in the TA response under pressure is likely an important reason for the reversal of thermal anisotropy, which may be related to the different changes in the group velocity.

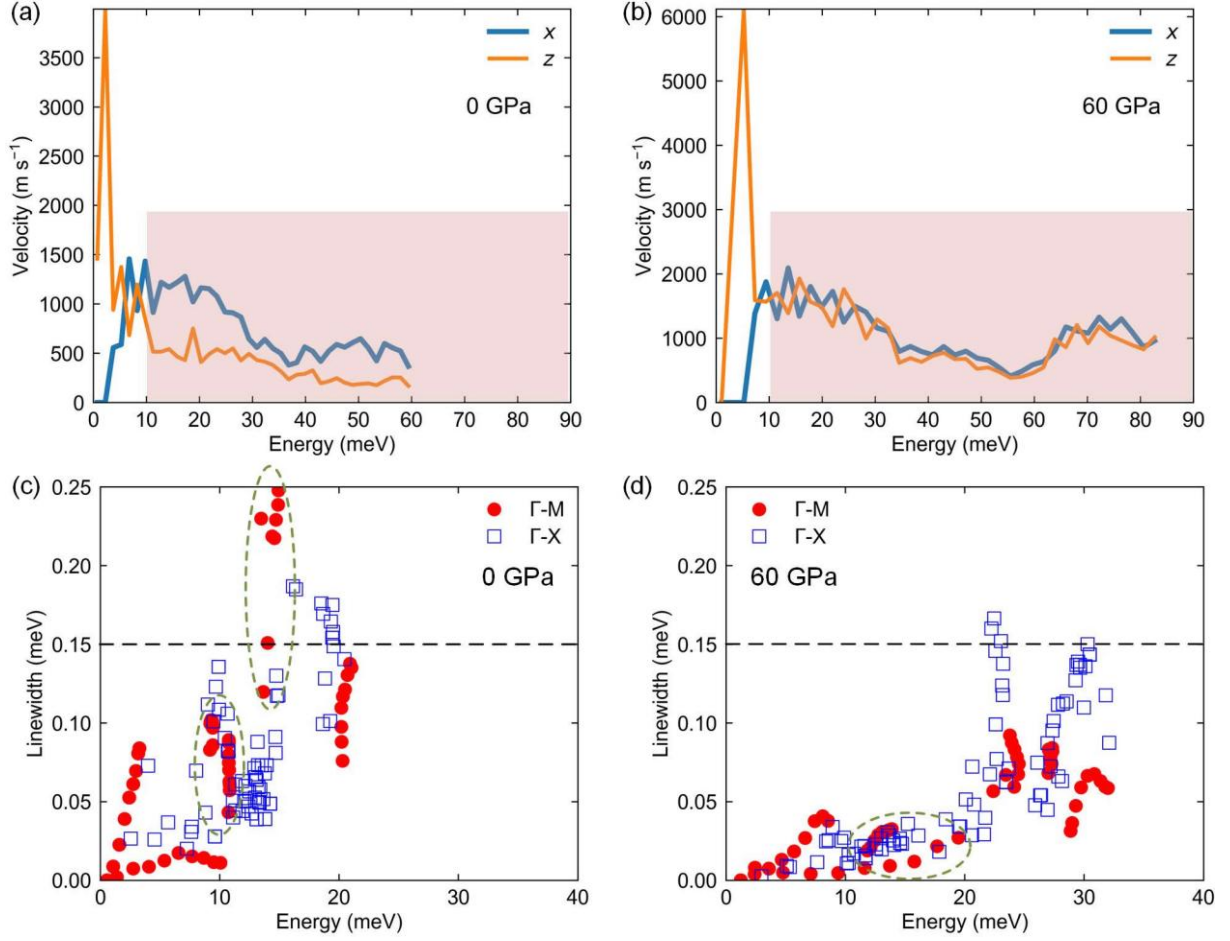

Figure 4 Phonon group velocities along the *z*- and *x*-directions as a function of phonon energy at (a) 0 GPa and (b) 60 GPa. Phonon linewidths along the Γ–M (*z*-direction) and Γ–X (*x*-direction) directions at (c) 0 GPa and (d) 60 GPa. The red dots within the dashed ellipses highlight the difference in linewidth distribution: at 0 GPa, the linewidths are concentrated near specific phonon energies, whereas they are spread over a broader energy range at 60 GPa.

To better understand the pressure-induced reversal of thermal conductivity anisotropy in $Bi_2O_2Se$, we examine the contributions from the heat capacity *C*, group velocity *v*, and phonon lifetime *τ*—the key components in the thermal conductivity expression $\kappa = \frac{1}{3}\sum_{\mathbf{q}s} C_{\mathbf{q}s} v_{\mathbf{q}s}^2 \tau_{\mathbf{q}s}$, where **q** and *s* denote the phonon wave vector and branch index, respectively. Because four-phonon scattering processes are negligible in this system, all analyses considered only three-phonon scattering processes. Notably, the heat capacity *C* is a scalar quantity and, thus, does not contribute to thermal anisotropy. We then focused on the group velocity and phonon lifetime.

Figure 4 shows the phonon group velocities along the *x*- and *z*-directions as a function of phonon energy at 0 and 60 GPa. The energy-dependent average group velocity is calculated by[25]

$$v(\omega) = \frac{(v^u)^2}{|v|}(\omega) = \left(\frac{\Delta q}{2\pi}\right)^3 \sum_\lambda \frac{(v_\lambda^u)^2}{v_\lambda^u}\delta(\omega - \omega_\lambda)\Big/\sum_j g_j(\omega)$$

where $v_\lambda^u$ denotes the group velocity of phonon mode $\lambda$ with wavevector **q** along the $\mu$-direction ($\mu$=x or z), and $g_j(\omega)$ represents the phonon density of states for branch $j$. At 0 GPa, the z-direction group velocity exceeds that along x below ~10 meV but becomes significantly lower at higher energies. In contrast, at 60 GPa, $v_z > v_x$ persists below 10 meV, and the two directions become nearly comparable above 10 meV (Figs. 4a and 4b). Notably, at 60 GPa, the magnitude of $v_z$ exceeding $v_x$ below 10 meV is significantly larger than that at 0 GPa.

Finally, the phonon lifetime was analyzed. Because phonons below 20 meV dominate the thermal transport (Fig. 3c), we focus on their linewidths ($\gamma$), which are inversely proportional to the lifetime ($\gamma = \tau^{-1}$). At 0 GPa (Fig. 4c), the phonon linewidths along Γ-M (z-direction) and Γ-X (x-direction) show no systematic difference; however, the linewidths of Γ-M modes cluster near specific energies (dashed ellipse in Fig. 4c), indicating quasi-flat bands with low group velocity and strong scattering—limiting $\kappa_z$. At 60 GPa (Fig. 4d), this clustering vanishes, where the Γ–M modes disperse over a broad energy range, consistent with the emergence of dispersive phonon branches and the convergence of $v_z$ and $v_x$ in the 10-20 meV range. Moreover, $\gamma$ along the z direction are generally slightly smaller than those along the x direction, suggesting longer lifetimes. This dispersion evolution reflects the disappearance of the low-energy "rattling-like" peak in the z-projected phonon DOS (Fig. 2c). Thus, the pressure-induced reversal of thermal anisotropy is primarily driven by the divergent evolution of group velocities along the Γ–M (z-direction) and Γ–X (x-direction) directions, with lifetime effects playing a secondary role. This behavior correlates with the structural changes under pressure. The z-axis contracts more than the x-axis (Fig. 1b), and the Bi atoms become less mobile along the z-direction (Fig. 1f). In addition, lone pair activity and the associated anharmonicity are more strongly suppressed along z (Figs. 2c and 2d). These effects lead to distinct stiffening of the phonon modes, especially the TA modes, ultimately resulting in a faster increase in $\kappa_z$ compared to $\kappa_x$.

In summary, our first-principles calculations combined with the phonon Boltzmann transport equation reveal that extreme hydrostatic pressure induces a reversal of thermal conductivity anisotropy in $Bi_2O_2Se$ without phase transition, that is, transformation from $\kappa_z < \kappa_x$ at ambient pressure to $\kappa_z > \kappa_x$ at high pressure. This reversal stems from anisotropic structural changes, in which the z-axis compresses more rapidly than the *x*-axis, leading to distinct phonon responses along the Γ–M (*z*) and Γ–X (*x*) directions. Notably, the low-energy "rattling-like" peak in the *z*-projected phonon DOS weakens under pressure, and the TA mode along Γ–M exhibits pronounced stiffening. These effects contribute to reshaping the group-velocity landscape. Above 10 meV, $v_x$ exceeds $v_z$ at 0 GPa, but this difference is significantly reduced at 60 GPa, where the two velocities become nearly comparable. Below 10 meV, $v_z$ surpasses $v_x$ at both pressures; however, the magnitude of this advantage is far larger at 60 GPa than at 0 GPa. This evolution underlies the accelerated increase in $\kappa_z$ relative to $\kappa_x$ under pressure and finally leads to the reversal of thermal conductivity anisotropy. Our findings challenge the conventional view that hydrostatic pressure universally suppresses thermal anisotropy toward isotropy, demonstrating instead that it can actively invert the direction of heat transport through anisotropic lattice responses—offering a potential route to pressure-driven solid-state thermal switches.

**Notes**

The authors declare no conflict of interest.

**Data Availability**

The data that supports the findings of this study are available within the article and its supplementary material.

**Acknowledgement**

This work is supported by Gansu Province Science and Technology Program Special Project of Soft Science (25JRZE003), Gansu Province Science and Technology Program (23ZDGE001), the National Science Foundation of China (Grant no 12304040) and GuangDong Basic and Applied Basic Research Foundation (Grant no 2023A1515110762). We thank the Beijing PARATERA Tech CO., Ltd. https://cloud.paratera.com.**Reference**

[1] B. Zhan, Y. Liu, X. Tan, J. Lan, Y. Lin, and C.-W. Nan, "Enhanced Thermoelectric Properties of Bi$_2$O$_2$Se Ceramics by Bi Deficiencies," J. Am. Ceram. Soc. **98**(8), 2465–2469 (2015).

[2] X. Ding, M. Li, P. Chen, Y. Zhao, M. Zhao, H. Leng, Y. Wang, S. Ali, F. Raziq, X. Wu, H. Xiao, X. Zu, Q. Wang, A. Vinu, J. Yi, and L. Qiao, "Bi$_2$O$_2$Se: A rising star for semiconductor devices," Matter **5**(12), 4274–4314 (2022).

[3] R. Guo, P. Jiang, T. Tu, S. Lee, B. Sun, H. Peng, and R. Yang, "Electrostatic interaction determines thermal conductivity anisotropy of Bi$_2$O$_2$Se," Cell Rep. Phys. Sci. **2**(11), (2021).

[4] R. Liu, J. Lan, X. Tan, Y. Liu, G. Ren, C. Liu, Z. Zhou, C. Nan, and Y. Lin, "Carrier concentration optimization for thermoelectric performance enhancement in *n*-type Bi$_2$O$_2$Se," J. Eur. Ceram. Soc. **38**(7), 2742–2746 (2018).

[5] D. Guo, C. Hu, Y. Xi, and K. Zhang, "Strain effects to optimize thermoelectric properties of doped Bi$_2$O$_2$Se via Tran–Blaha modified Becke–Johnson density functional theory," J. Phys. Chem. C **117**(41), 21597–21602 (2013).

[6] W.R. Lee, and C. Lee, "Enhancing the thermoelectric properties of layered Bi$_2$O$_2$Q (Q = S, Se): the effect of mixed chalcogen net," J. Korean Phys. Soc. **73**(11), 1684–1689 (2018).

[7] A. Novitskii, M.Y. Toriyama, I. Serhiienko, T. Mori, G.J. Snyder, and P. Gorai, "Defect engineering of Bi$_2$SeO$_2$ thermoelectrics," Adv. Funct. Mater. **35**(10), 2416509 (2025).

[8] A.L.J. Pereira, D. Santamaría-Pérez, J. Ruiz-Fuertes, F.J. Manjón, V.P. Cuenca-Gotor, R. Vilaplana, O. Gomis, C. Popescu, A. Muñoz, P. Rodríguez-Hernández, A. Segura, L. Gracia, A. Beltrán, P. Ruleova, C. Drasar, and J.A. Sans, "Experimental and theoretical study of Bi$_2$O$_2$Se under compression," J. Phys. Chem. C **122**(16), 8853–8867 (2018).

[9] Y. Zhou, Z.-Y. Dong, W.-P. Hsieh, A.F. Goncharov, and X.-J. Chen, "Thermal conductivity of materials under pressure," Nat. Rev. Phys. **4**(5), 319–335 (2022).

[10] L. Lindsay, D.A. Broido, J. Carrete, N. Mingo, and T.L. Reinecke, "Anomalous pressure dependence of thermal conductivities of large mass ratio compounds," Phys. Rev. B **91**(12), 121202 (2015).

[11] Y. Li, J. Liu, X. Wang, and J. Hong, "Anomalous suppressed thermal conductivity in CuInTe$_2$ under pressure," Appl. Phys. Lett. **119**(24), 243901 (2021).

[12] H. Yu, L.-C. Chen, H.-J. Pang, X.-Y. Qin, P.-F. Qiu, X. Shi, L.-D. Chen, and X.-J. Chen, "Large enhancement of thermoelectric performance in CuInTe$_2$ upon compression," Mater. Today Phys. **5**, 1–6 (2018).

[13] L.-C. Chen, H. Yu, H.-J. Pang, B.-B. Jiang, L. Su, X. Shi, L.-D. Chen, and X.-J. Chen, "Pressure-induced enhancement of thermoelectric performance in palladium sulfide," Mater. Today Phys. **5**, 64–71 (2018).

[14] N.K. Ravichandran, and D. Broido, "Non-monotonic pressure dependence of the thermal conductivity of boron arsenide," Nat. Commun. **10**(1), 827 (2019).

[15] K. Ohta, Y. Nishihara, Y. Sato, K. Hirose, T. Yagi, S.I. Kawaguchi, N. Hirao, and Y. Ohishi, "An experimental examination of thermal conductivity anisotropy in *hcp* iron," Front. Earth Sci. **6**, (2018).14